\documentclass[aps,prd,twocolumn,fleqn,superscriptaddress]{revtex4}
\usepackage{graphicx,color,natbib}
\usepackage{amsmath,amssymb,amsfonts}

\newcommand{\bse}{\begin{subequations}}
\newcommand{\ese}{\end{subequations}}
\newcommand{\be}{\begin{equation}}
\newcommand{\ee}{\end{equation}}
\newcommand{\bea}{\begin{eqnarray}}
\newcommand{\eea}{\end{eqnarray}}
\newcommand{\ba}{\begin{array}}
\newcommand{\ea}{\end{array}}

\input amssym.def
\input amssym.tex

\usepackage[colorlinks=true, linkcolor=blue, bookmarks=true]{hyperref}

\begin{document}

\title{AdS/QCD, Entanglement Entropy and Critical Temperature}

\author{M. Lezgi\footnote{mahsalezgee@yahoo.com}}
\affiliation{Department of Physics, Shahid Beheshti University G.C., Evin, Tehran 19839, Iran}
\author{M. Ali-Akbari\footnote{$\rm{m}_{-}$aliakbari@sbu.ac.ir}}
\affiliation{Department of Physics, Shahid Beheshti University G.C., Evin, Tehran 19839, Iran}

\begin{abstract}
Based on gauge-gravity duality, by using holographic entanglement entropy, we have done a phenomenological study to probe confinement-deconfinement phase transition in the QCD-like gauge theory.
Our outcomes are in perfect agreement with the expected results, qualitatively and quantitatively. We find out that the (holographic) entanglement entropy is a reliable order parameter for probing the phase transition.
\end{abstract}

\maketitle


\section{Introduction}
In the quantum chromodynamics (QCD), explaining confinement-deconfinement phase transition is one of the most interesting and challenging issues. It is experimentally expected that this phase transition takes place at low temperatures, in the region of about $175\pm 10$ MeV \cite{Satz:2011wf}. Since the coupling of the QCD is large at low temperatures, it is impossible to describe this phase transition by using usual perturbative methods, correctly. Therefore, we need a non-perturbative framework to study the phase transition and, for instance, answer the following significant questions: What is the relevant order parameter of the phase transition? At which temperature, depending on the parameters of the theory, does the transition occur? At which distance is the potential between a quark and antiquark screened? How is the binding energy of the quark and antiquark determined and calculated?  

Gauge-gravity duality (or holography idea), as a non-perturbative method, introduces a new link between gauge theory and gravity. More precisely, according to this duality, a strongly coupled quantum gauge theory defined in a $d$-dimensional space-time corresponds to a classical gravity in a $d+1$-dimensional space-time \cite{CasalderreySolana:2011us, Maldacena, Natsuume:2014sfa}. Therefore, different questions in the strongly coupled gauge theory can be translated into corresponding problems in the classical gravity. Consequently, this idea has been applied to optimize ansatz for the gravity dual, perhaps with unknown gauge theory dual, in such a way that the known features of QCD can be reproduced. This is usually called the AdS/QCD approach \cite{ahmad}. Applying AdS/QCD, the mentioned questions have been extensively discussed in the literature, most of them qualitatively. For instance see \cite{CasalderreySolana:2011us} and references therein.  

Gauge-gravity duality is also introduced as a useful prescription to calculate entanglement entropy, which is one of the interesting physical quantities on the gauge theory side \cite{Nishioka:2009un}. Then, interestingly in \cite{Klebanov:2007ws},  the entanglement entropy is used to probe a confinement-deconfinement phase transition at zero temperature in confining theories. Hence, search for transition has also been extended to (non-conformal) gauge theories at finite temperature \cite{Faraggi:2007fu, Rahimi:2016bbv, Knaute:2017lll}. It is shown that no transition takes place at finite temperature. In this paper, we are going to answer the mentioned remarkable questions in the context of gauge-gravity duality, or more precisely AdS/QCD,  by using the holographic entanglement entropy as an order parameter of the confinement-deconfinement phase transition.

\section{Review on Model, potential energy and entanglement entropy}
We consider the following background metric
\be\label{metric} %
 ds^2=\frac{R^2}{z^2}g(z)\left(dt^2+d\vec{x}^2+dz^2\right)
\ee %
where $g(z)=e^{\frac{c}{2}z^2}$ and $\vec{x}\equiv (x_1, x_2, x_3)$. The radial direction is denoted by $z$. According to holography idea, the QCD-like gauge theory is living on the boundary of the above background which is located at $z=0$. As it is clearly seen the metric \eqref{metric} approaches AdS$_5$ with radius $R$, asymptotically. In order to fit the result of the above metric with the slope of Regge trajectories, it is shown that $c\approx 0.9$ GeV$^2$ \cite{Andreev:2006vy}. Various properties of this background have been investigated in \cite{Andreev:2006ct, Andreev:2006eh}.

\textit{Potential energy:} In the gauge theory, the static potential energy between a quark and an anti-quark is evaluated by using the expectation value of the Wilson loop on a rectangular loop, ${\cal{C}}$, that contains two sides, time $\cal{T}$ and distance $r$, where the length of time direction is much larger than the distance between the quarks , i.e. ${\cal{T}}\gg r$. Then it is easy to show that \cite{CasalderreySolana:2011us}
\be\label{staticwilson}
\langle W({\cal{C}}) \rangle  =e^{-i(2 m + V(r)){\cal{T}}},
\ee%
where $m$ is the rest mass of the quarks and $V(r)$ is the potential energy between them. 
The holographic dual of the rectangular Wilson loop is given by a classical string suspended from two points on the boundary (corresponding to quark and anti-quark), hanging down in extra dimension with appropriate boundary conditions \cite{Maldacena:1998im}. 	In fact, the expectation value of the Wilson loop is dual to the on-shell action of a string the endpoints of which are separated by a distance $r$ \cite{CasalderreySolana:2011us, Maldacena:1998im}. 


The dynamics of a classical string in an arbitrary background is described by 
\be %
\label{action}
 S=\frac{-1}{2 \pi \alpha'} \int d\tau d\sigma  \sqrt{- \det(g_{ab})}.
\ee %
where $g_{ab}$ is the induced metric on the world-sheet and is defined by $g_{ab}=G_{MN}\frac{\partial X^M}{\partial \xi^a} \frac{\partial X^N}{\partial \xi^b}$. $X^M$ ($\xi^a=\tau,\ \sigma$) denotes the space-time (world-sheet) coordinates and $G_{MN}$ is the background metric.
Thus, according to gauge-gravity duality, we have
\be\label{wilson}
\langle W({\cal{C}})\rangle =e^{i S({\cal{C}})}.
\ee%
We will now proceed to calculate $S({\cal{C}})$ for the rectangular loop ${\cal{C}}$ to find the potential energy from \eqref{staticwilson}. For the metric \eqref{metric}, all calculations have been done in detail and with the ansatz $t=\tau$, $x_1=\sigma$ and $z(x_1)$, the result for distance $r$ is given by \cite{Andreev:2006ct}
\be\begin{split}
r=2\sqrt{\frac{\lambda}{c}}\int_{0}^{1}dvv^2e^{\frac{1}{2}\lambda\left(1-v^2\right)}h(v,\lambda),
\end{split}\ee %
where $h(v,\lambda)=[1-v^4e^{\lambda\left(1-v^2\right)}]^{-\frac{1}{2}}$, $v=z/z_*$, $\lambda=cz_*^2$ and $z_*=z(x_1=0)$ which is the turning point of the string. It is easy to check that the distance $r$ is real for $0<\lambda<2$ corresponding to $0<r<\infty$. In other words, there is an upper limit on the turning point of the string, that is $z_*<\sqrt{2/c}$.
Note that for the case of the AdS$_5$, i.e. $c=0$, there is no bound on the $z_*$, as expected. Then the potential energy in the QCD-like gauge theory is also given by \cite{Andreev:2006ct}
\be\label{potential} %
V(r)=\frac{p}{\pi}\sqrt{\frac{c}{\lambda}} 
\left(\int_{0}^{1}dvv^{-2}\left[e^{\frac{1}{2}\lambda{v^2}}
h(v,\lambda)-1\right]-1\right),
\ee %
where its behaviour at large and short distance can be found as follows
\bea\label{maximum} %
 V(r)=\left\{%
\begin{array}{ll} %
    p\left(-\frac{\kappa_0}{r}+\sigma_0 r+ O(r^3)\right), \ \ \ \ \ r\rightarrow 0 \\
    p(\sigma r), \ \ \ \ \ \ \ \ \ \ \ \ \ \ \ \ \ \ \ \ \ \ \ \ \ \ \ \ r\rightarrow\infty\\
\end{array}%
\right.
\eea %
where $p\approx0.94$, $\kappa_0\approx0.23$, $\sigma_0\approx0.16$ GeV$^2$ and $\sigma\approx0.19$ GeV$^2$ for $c=0.9$ GeV$^2$ \cite{Andreev:2006ct}. This potential gives the expected linear and $1/r$ behaviour at large and short distance.

\textit{Entanglement Entropy:} we can take into account a quantum system with many degrees of freedom at zero temperature. This system is described by a pure ground state $| \psi \rangle$ and subsequently its density matrix is determined by $\rho=| \psi \rangle\langle \psi |$.  This quantum system may be divided into two subsystems $A$ and $B$, with the observer in the subsystem $A$ not having access to the degrees of freedom of subsystem $B$. Therefore the system's density matrix can be determined by taking the trace over these degrees of freedom, i.e. $\rho_A=tr_B\rho$. Thus subsystem $A$'s entanglement entropy will be $S_A=-tr_A(\rho_A\log\rho_A$) which shows the amount of information lost when an observer is limited to the subsystem $A$.  In the case of a gauge theory in $d>2$ space-time, the main divergence of $S_A$ is proportional to the area of the subsystem $A$.  In the case of a two-dimensional gauge theory, in which subsystem $A$ is an interval of length $l$, we can analytically calculate the entanglement entropy as a universal result $S_l=\frac{c}{3}\log(\frac{l}{a})$ where $c$ is central charge and $a$ is the UV cut-off of the field theory.   
     
On the holographic side, entanglement entropy $S_A$ may be calculated through the following formula \cite{Nishioka:2009un}
\be %
 S_A=\frac{\rm{Area}(\gamma_A)}{4G_5},
\ee %
where $\gamma_A$ is a three-dimensional minimal area surface in asymptomatically $AdS_5$ background whose boundary is given by $\partial A$ (which is the boundary of the subsystem $A$). This simple prescription leads to well-known results, such as entanglement entropy in two-dimensional conformal field theory, thus it is reliable to calculate the entanglement entropy in the strongly coupled gauge theories.  

We begin with a general form for the background as
\be\label{metric1}
 ds^2=-f_1(z)dt^2+f_2(z) dz^2+f_3(z)d\vec{x}^2 ,
\ee 
where $z$ is the radial direction. The background is asymptotically $AdS_5$ and its boundary is located at $z=0$. We must divide the boundary region into two subsystems $A$ and $B$ in order to determine entanglement entropy. Subsystem $B$ is defined by $-\frac{l}{2}<x_1(\equiv x)<\frac{l}{2}$ and $x_2, x_3\in(-\infty,+\infty)$ at a given time. Then the minimal area of $\gamma_A$, which is proportional to entanglement entropy of subsystem $A$, is obtained by minimizing the following area 
\be\label{area} %
 S_A^{(c)}=\frac{1}{4G_5}\int d^3 x \sqrt{g_{in}},
\ee %
where $G_5$ is the five-dimensional Newton constant and $g_{in}$ is induced metric on $\gamma_A$. Using \eqref{metric1}, we can easily see that  
\be\label{EE1}
S_A^{(c)}=\frac{V_2}{4G_5}\int_{\frac{-l}{2}}^{\frac{l}{2}} dx \sqrt{f_3^3(z)+f_3^2(z) f_2(z) z^{\prime 2}
},
\ee
where $V_2$ is the area of two-dimensional surface of $x_2$ and $x_3$ and $z'=\frac{dz}{dx}$. The above area is not explicitly dependent on $x$ so the   corresponding Hamiltonian is a constant of motion
\begin{align}\label{derevative}
\frac{f_3^2\left(z\right)}{\sqrt{f_3\left(z\right)+f_2\left(z\right)z^{\prime 2}}}={\rm{const}}=f_3^\frac{3}{2}(z_*),
\end{align}
where $z_*$ is the minimal value of $z$, i.e. $z(x=0)=z_*$, and $z'(x=0)=0$. Hence, from \eqref{derevative}, we get to 
\begin{align}\label{zprime}
z^\prime=\sqrt{\frac{f_3(z)}{f_2(z)}}\sqrt{\frac{f_3^3(z)}{f^3_3(z_*)}-1},
\end{align}
and then we can easily obtain the relation between $l$ and $z_*$  
\begin{align}
l=2\int_0^{z_*}\sqrt{\frac{f_2(z)}{f_3(z)}}\frac{dz}{\sqrt{\frac{f_3^3(z)}{f_3^3(z_*)}-1}}.
\end{align}
Finally, by substituting \eqref{zprime} in \eqref{EE1}, we have 
\be\label{entanglement}
S_A^{(c)}=\frac{V_2}{2G_5}\int_{0}^{z_*}\frac{f_3^\frac{5}{2}\left(z\right)f_2^{\frac{1}{2}}\left(z\right)}{\sqrt{f_3^3\left(z\right)-f_3^3\left(z_*\right)
}}dz.
\ee
Since the factor behind the integral in \eqref{entanglement} is obviously a constant we just need to compute the integral, which is proportional to the entanglement entropy.       

Another configuration we would like to consider here is disconnected solution. This configuration is described by two disconnected surfaces located at $x=\pm l/2$ and extended in all other spatial directions. It is easy then to calculate the entanglement entropy as 
\be 
S_A^{(d)}=\frac{V_2}{2G_5}\int_{0}^{\sqrt{\frac{2}{c}}}f_3(z)f_2^{\frac{1}{2}}\left(z\right)dz,
\ee 
and we then introduce
\be 
\Delta S(l)\equiv\frac{2G_5}{V_2}\left(S_A^{(c)}-S_A^{(d)}\right).
\ee %
Now, one can define the critical distance $l_c$ as $\Delta S(l_c)=0$, i.e. the length at which the difference between entanglement of the connected and disconnected surfaces is zero. Then for $l>l_c$ ($l<l_c$), corresponding to $\Delta S>0$ ($\Delta S<0$), the entanglement scales as $N_c^0$ ($N_c^2$) where $N_c$ is the number of colors in the gauge theory \cite{Klebanov:2007ws}. This change of behaviour at $l=l_c$ resembles a confinement-deconfinement first order phase transition in the dual gauge theory.

\section{Numerical results}
\begin{figure}[ht]
\begin{center}
\includegraphics[width=72 mm]{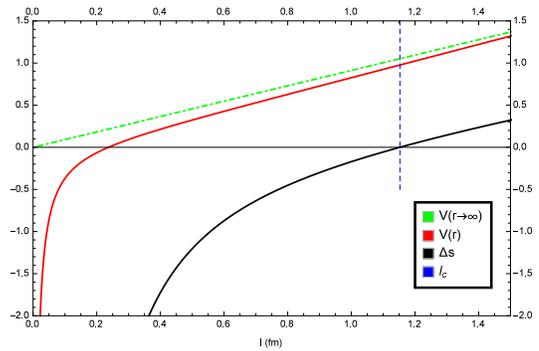}
\caption{ The difference of the entanglement entropies $\Delta S$, potential $V(r)$ and asymptotic potential $V(r\rightarrow\infty)$ as a function of $l$ for $c=0.9$ GeV$^2$. \label{various}}
\end{center}
\end{figure}

At zero temperature, according to \eqref{metric}, it is clear that $f_1(z)=f_2(z)=f_3(z)=\frac{R^2}{z^2}g(z)$. In figure \ref{various}, we have plotted the difference of the entanglement entropies $\Delta S$, potential $V(r)$ and asymptotic potential $V(r\rightarrow\infty)$ as a function of $l(=r)$. Since we are working with the confining background \eqref{metric}, the potential energy \eqref{potential} always describes a stable quark-antiquark bound state in the confined phase, although its sign changes in small values of $l$. However, the difference of the entanglement entropies $\Delta S$ predicts that for $l<l_c$ the system under study is not in the confined phase anymore and as a result a phase transition between confined and deconfined phases occurs. It is worthy to note that the phase transition happens in the regime of $l$ for which the linear behaviour of the potential has been dominated, as expected. Another remarkable result is that the phase transition occurs at $l_c\approx 1$ fm for the reasonable values of $c$, as we will discuss. In other words, the allowed quark-antiquark separation is about 1 fm, compatible with the size of the hadrons (which is also 1 fm).
\begin{figure}[ht]
\begin{center}
\includegraphics[width=72 mm]{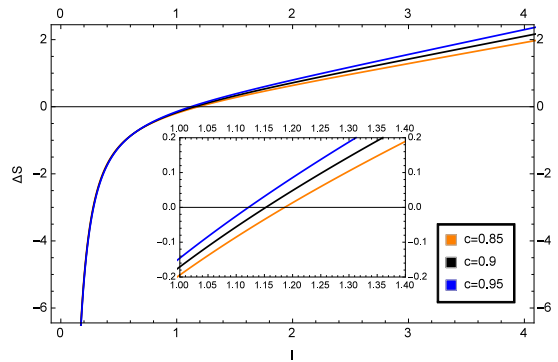}
\caption{ $\Delta S$ in terms of $l$ for three diffrenet values of $c$. \label{eelnew}}
\end{center}
\end{figure}

Let us return to the reasonable values for $c$. As it was stated, $c\approx 0.9$ GeV$^2$ has been used in various papers. In figure \ref{eelnew}, $\Delta S$ has been plotted in terms of distance $l$ for three different values of $c$. This figure shows that by increasing $c$, the critical length $l_c$ decreases. Although this reduction is not substantial, it plays a very important role in determining the critical temperature $T_c\simeq l_c^{-1}$. In fact, $c$ and critical temperature increase together, as it is easily seen in table \ref{list}.

\begin{figure}[ht]
\begin{center}
\includegraphics[width=72 mm]{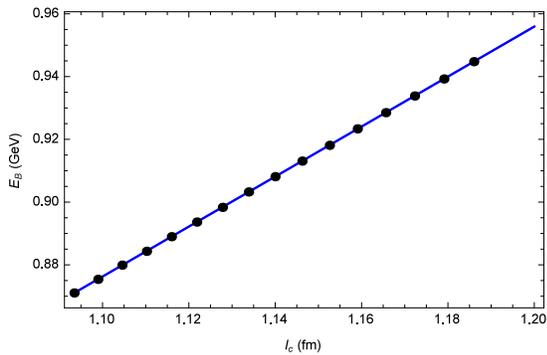}
\caption{ The binding energy in terms of critical length. The black points are numerical results and the fitted line is blue.  \label{binding}}
\end{center}
\end{figure}

As an another interesting result, one can estimate the value of the binding energy between a quark and an antiquark. One can define the binding energy as $E_B\simeq V(l_c)$ where the critical length $l_c$ denotes the maximum size of the bound state. Figure \ref{binding} shows that the binding energy linearly increases by $l_c$ for $0.85<c<0.95$. According to our results, the value of the binding energy alters between 0.87 and 0.96 GeV$^2$ which is in agreement with the expected result, $0.5<E_B<1$ GeV$^2$ \cite{Satz:2011wf}. The black points in figure \ref{binding} are fitted with $E_B=0.796608\  l_c$ (blue line).

Now let us consider the following background metric 
\be\label{metric12} %
 ds^2=\frac{R^2}{z^2}g(z)\left(f(z)dt^2+d\vec{x}^2+\frac{dz^2}{f(z)}\right)
\ee %
where $f(z)=1-z^4/z_h^4$. The horizon is denoted by $z_h$ and therefore the Hawking temperature is given by $T=\frac{1}{\pi z_h}$. Then it was discussed in \cite{Andreev:2006eh} that a confinemet-deconfinemet phase transition takes place when $z_h=z_c=\sqrt{2/c}$ or equivalently $T_c^*=\frac{1}{\pi}\sqrt{\frac{c}{2}}$. In other words, on the gravity side the phase transition is described by changing the background geometry from \eqref{metric} to \eqref{metric12}.  The values of $T_c^*$ have been shown in the table \ref{list} for reasonable values of $c$. It is clearly seen that the critical temperature can be better described by our results since $T_c$ is about 175$\pm$10 MeV.
\begin{table}[ht]
\caption{Critical temperatures (MeV) and length (fm)}
\vspace{1 mm}
\centering
\begin{tabular}{c c c c c  c c c}
\hline\hline
$c$ & 0.86 & 0.88 & 0.90 & 0.92 & 0.94 & 0.96 & 0.98\\
\hline\hline
$l_c$ & 1.17905 &  1.16558  & 1.15255 & 1.13995 & 1.12776 & 1.11595 & 1.1045\\
$T_c$ & 167.30 &  169.27 & 171.18 & 173.13 & 174.95 & 176.80 & 178.63\\
$T_c^*$ & 208.50 & 210.91  & 213.30 & 215.65 & 217.98 & 220.29 & 222.57\\
\hline
\end{tabular}\\[1ex]
\label{list}
\end{table}

Table \ref{list} also indicates that in order to find better values for the critical temperature and length, the parameter $c$ should be estimated more precisely. Our results, by applying the entanglement entropy as a probe for the confinment-deconfinement phase transition, predict that the sensible value for the parameter $c$ is in the range between 0.85 and 1.1 GeV$^2$. However, our calculation indicates that $c=0.94$ GeV$^2$ is a better choice.

An extension of the above idea to the thermal backgrounds, introduced in \cite{Klebanov:2007ws}, shows that entanglement entropy does not exhibit a phase transition (expect for geometry of the near horizon limit of D6-branes) \cite{Faraggi:2007fu}. Therefore, as a final point, we would like to investigate the possibility of a phase transition in the presence of an event horizon corresponding to a thermal field theory. To do so, two types of surfaces have been considered: connected and piecewise smooth. The connected surface has been already discussed in \eqref{entanglement} for an arbitrary background. The second surface is defined as 
\be %
 x=-\frac{l}{2},\ \ \ \ \ z=z_h,\ \ \  x=\frac{l}{2},
\ee %
and then, using \eqref{area}, one gets
\begin{align}\label{ss}
\hat{S}^{(d)}_A=\frac{V_2}{4G_5}\left(2\int_{0}^{z_h} dz f_3(z)\sqrt{f_2(z)}+l\sqrt{f_3^3(z_h)}\right),
\end{align}
where $f_3=\frac{R^2}{z^2}g(z)$ and $f_2=\frac{R^2g(z)}{z^2f(z)}$. By defining the following expression 
\be %
 \hat{\Delta} S\equiv\frac{2G_5}{V_2}\left(S^{(c)}_A-\hat{S}^{(d)}_A\right),
\ee %
we observe that, in our case, phase transition does not take place at finite temperature. In other words, $\hat{\Delta} S$ is always negative and as a result the connected surface is favourable for all values of $l$.

\section*{Acknowledgement}
We would like to thank School of Physics of Institute
for research in fundamental sciences (IPM) for the research facilities and
environment. The authors would also like to thank H. Ebrahim and M. M. Sheikh Jabbari for useful
comments.

\end{document}